\documentstyle[sprocl]{article}
\def\Journal#1#2#3#4{{#1} {\bf #2}, #3 (#4)}

\def\be{\begin{equation}}
\def\ee{\end{equation}}
\def\bea{\begin{eqnarray}}
\def\eea{\end{eqnarray}}

\def\AnnP{\em Ann. Phys.}
\def\NP{\em Nucl. Phys.}
\def\PL{\em Phys. Lett.}
\def\PR{\em Phys. Rev.}
\begin{document}
\begin{flushright}
hep-ph/9609386 \\
TTP96-46 \\
YCTP-P18-96 \\
August 1996
\end{flushright}
\vspace*{1cm}
\title{EFFECTIVE FIELD THEORY FOR A HEAVY HIGGS BOSON:
MATCHING AND GAUGE INVARIANCE
\footnote{Talk presented by A.N.\ at the DPF96 Conference, Minneapolis,
MN, August~10 --~15, 1996, to appear in the Proceedings.}}
\author{ A. NYFFELER }
\address{Yale University, Dept.\  of Physics, New Haven,
CT 06520-8120, USA \\
nyffeler@genesis2.physics.yale.edu}
\author{ A. SCHENK }
\address{Univ.\  Karls\-ruhe, Inst.\  f\"ur Theor.\  Teilchenphysik,
D-76128 Karlsruhe, Germany \\
andreas.schenk@phys.uni-karlsruhe.de}
%
\maketitle\abstracts{
For large values of the Higgs mass the low energy structure of the
gauged linear sigma model in the spontaneously broken phase can
adequately be described by an effective field theory. We present a
manifestly gauge-invariant functional technique to explicitly evaluate
the corresponding effective Lagrangian from the underlying theory.
}
\section{Introduction}

The method of effective field theory has repeatedly been used in the
analysis of the symmetry breaking sector of the Standard Model. It
provides a model independent parametrization of various scenarios for
the spontaneous breakdown of the electroweak symmetry. The unknown
physics is then hidden in the low energy constants of an effective
Lagrangian.

The physics in the low energy region of a full theory is adequately
described by an effective field theory if corresponding Green
functions in both theories have the same low energy structure. One can
take this matching requirement as the definition of the effective
field theory. It determines functional relationships between the low
energy constants of the effective Lagrangian and the parameters of the
underlying theory.

The special role of gauge theories is readily understood. The
effective field theory analysis should not make any particular
assumptions about the underlying theory -- apart from symmetry
properties and the existence of a mass gap.  This also requires
parametrizing low energy phenomenology by a gauge-invariant effective
Lagrangian. The definition of Green's functions, on the other hand,
usually does not reflect the symmetry properties of a gauge theory.
Gauge invariance is broken, and the off-shell behaviour of Green's
functions is gauge-dependent. Therefore, if the Green functions
which enter the matching relations do not reflect the symmetry
properties of the full theory, the effective field theory will also
include the corresponding gauge artifacts.

Without resolving these issues, different approaches to determine the
low energy constants at order $p^4$ for the Standard Model with a
heavy Higgs boson where presented in several recent
articles~\cite{SM_Heavy_Higgs}. For Higgs masses below about~$1 TeV$
the effective Lagrangian can be evaluted explicitly with perturbative
methods. These works show clearly that the matching of gauge-dependent
quantities causes all kinds of trouble. This calls for a new
manifestly gauge-invariant technique which avoids these problems, yet
maintains the simplicity and elegance of matching Green's functions as
in the ungauged case.

\section{A gauge-invariant approach}

Any approach to determine the effective Lagrangian for a given
underlying gauge theory should match only gauge-invariant quantities.
Then one does not have to worry about any gauge artifacts which
otherwise might enter the effective field theory. Perhaps the most
straightforward idea that comes to mind is to match only $S$-matrix
elements. However, this approach is quite cumbersome. In particular,
it involves a detailed treatment of infrared physics.

Functional techniques like those described in Ref.~\cite{LSM} provide
a much easier approach. In this case one matches the generating
functionals of Green's functions in the full and the effective theory.
Infrared physics drops out at a very early step of the calculation.
The remaining contributions all involve the propagation of heavy
particles over short distances. Hence, they can be evaluated with a
short distance expansion. The computation of loop-integrals is not
necessary.

Gauge invariance is broken as soon as Green's functions of
gauge-depend\-ent operators are considered. Hence, any manifestly
gauge-invariant approach must confine itself to analyze Green's
functions of gauge-invariant operators, such as the field strength of
an Abelian gauge field or the density of the Higgs field. In the
following we summarize a manifestly gauge-invariant technique to
evaluate the effective Lagrangian describing the low-energy region of
the gauged linear sigma model in the spontaneously broken phase. It
involves only Green's functions of gauge-invariant operators. For any
details the reader is referred to Ref.~\cite{Abelian} where it has
been applied to the Abe\-lian case. The Higgs sector of the Standard
model with the non-abelian group $SU_L(2)\times U_Y(1)$ can be treated
in the same way~\cite{NonAbelian}. We would like to point out, that
all $S$-matrix elements of the theory can be evaluated from these
Green functions as well.

At tree-level the generating functional is given by the classical
action. Since the external sources are gauge invariant, i.e., do not
couple to the gauge degrees of freedom, the equations of motion can be
solved without gauge-fixing. As a result they have a whole class of
solutions. Every two representatives are related to each other by a
gauge transformation. In order to determine the leading contributions
to the low energy constants one merely has to solve the equation of
motion for the Higgs boson field.

To incorporate higher order corrections one may evaluate the
path-integral representation of the generating functional with the
method of steepest descent. In this case they are described by
Gaussian integrals. Since gauge invariance is manifest, these
integrals can also be evaluated without gauge-fixing. As a consequence
the gauge degrees of freedom manifest themselves through zero-modes of
the quadratic form in the exponent of the gaussian factor. The
integration over these modes yields the volume of the gauge group,
which can be absorbed by the normalization of the path-integral. The
remaining integral over the non-zero modes contains all the physics.

We would like to point out one difference between this approach to
evaluate a path-integral and the method of Faddeev and Popov: if the
gauge is not fixed the evaluation of the path-integral does not involve
ghost fields. Hence, the number of diagrams to compute is reduced.

The effective Lagrangian of the linear sigma model is a sum of
gauge-invariant terms with an increasing number of covariant
derivatives and gauge-boson mass factors, corresponding to an
expansion in powers of the momentum and the masses. Note that the
covariant derivative, the gauge-boson masses and the gauge couplings
all count as quantities of order $p$. Thus, the low energy expansion
is carried out such that all light-particle singularities are
correctly reproduced. Furthermore, if the coupling $\lambda$ of the
scalar field is small enough, the low energy constants in the
effective Lagrangian admit an expansion in powers of this quantity,
corresponding to the loop expansion in the full theory. $n$-loop
Feynman diagrams in the Abelian Higgs model yield corrections of order
$\lambda^{n-1}$ to the low energy constants.

\newpage
\section*{Acknowledgments}
Work supported in part by Deutsche Forschungsgemeinschaft Grant Nr.\
Ma 1187/7-2, DOE grant \#DE-FG02-92-ER40704, NSF grant PHY-92-18167,
and by Schweizerischer Nationalfonds.
%
%

%
\end{document}